\documentclass[preprint,prd,aps,showpacs,showkeys,nofootinbib]{revtex4}
\usepackage{amssymb}
\usepackage{amsmath}

\usepackage{graphicx}
\usepackage{subfigure}

\usepackage{fancyhdr}

\usepackage{float}

\usepackage{fancyhdr}

\begin{document}

\preprint{Submitted to 'Chinese Physics C'}

\title{Top quark decay to a $125\;{\rm GeV}$ Higgs in BLMSSM}

\author{Tie-Jun Gao$^{a,b,c}$, Tai-Fu Feng$^{a,b}$, Fei Sun$^{a,c}$, Hai-Bin Zhang$^{a,c}$, Shu-Min Zhao$^{a,b}$ }

\affiliation{$^a$Department of Physics, Hebei University, Baoding, 071002, China\\
$^b$Institute of theoretical Physics, Chinese Academy of Sciences,
Beijing, 100190, China\\
$^c$Department of Physics, Dalian University of
Technology, Dalian,
116024, China}

\begin{abstract}
In this paper, we calculate the top quark rare decay $t\rightarrow
ch$  in a supersymmetric extension of the standard model where
baryon and lepton numbers are local gauge symmetries. Adopting
reasonable assumptions on the parameter space, we find that the
branching ratios of $t\rightarrow ch$ can reach $10^{-3}$, which can
be detected in near future.

\end{abstract}


 \maketitle

\section{Introduction\label{sec1}}
\indent\indent

Top quark plays a special role in the standard model(SM) and holds
great promise in revealing the secret of new physics beyond the SM.
The running LHC is a top-quark factory, and provides a great
opportunity to seek out top-quark rare decays.  Among those rare
processes, the flavor-changing neutral current (FCNC) decays
$t\rightarrow ch$ deserve special attention, since the branching
ratios(BRs) of those rare processes are strongly suppressed in the
SM. In addition, ATLAS and CMS have reported significant excess
events which are interpreted probably to be related to the neutral
Higgs with mass $m_{h_0}\sim 124-126\;{\rm GeV}$\cite{ref1,ref2}.
This implies that the Higgs mechanism to break electroweak symmetry
possibly has a solid experimental cornerstone.

In the framework of the SM, the possibility of detecting FCNC decays
$t\rightarrow ch$  is essentially hopeless, since tree level FCNC involving the quarks are forbidden by the gauge symmetries and particle content\cite{ref201,ref202}. In particular, it has
recently been recognized that the BRs of the process is much smaller
\cite{ref3,ref4} than originally thought \cite{ref5}  which is less
then $10^{-13}$. In extensions of the SM, the BRs for FCNC top
decays can be orders of magnitude larger. For example, the authors
of Ref.\cite{ref6}, study $t\rightarrow ch$ process  in the
framework of the minimal supersymmetric extension of the standard
model(MSSM) include the leading set of supersymmetric QCD and
supersymmetric electroweak contributions, and get Br$^{SUSY-EW}(t
\rightarrow ch) \sim10^{-8}$, Br$^{SUSY-QCD}(t \rightarrow ch)
\sim10^{-5}$. And a new work about this process  in MSSM is discussed in Ref.\cite{ref61}, with $\tan\beta=1.5\ or\ 35$ and the mass of SUSY particles about 1 or 2TeV scale, the authors get the branching ratio of $t \rightarrow ch$
can only reach $3\times10^{-6}$,  which is much smaller than previous results obtained before the advent of the LHC.

Physicists have been interested in the MSSM
\cite{ref7,ref71,ref72,ref73} for a long time. However, since the
matter-antimatter is asymmetry in the universe, baryon number (B)
should be broken. On the other hand, since heavy majorana neutrinos
contained in the seesaw mechanism can induce the tiny neutrino
masses\cite{ref8,ref81} to explain the neutrino oscillation
experiment, so the lepton number (L) is also expected to be broken.
A minimal supersymmetric extension of the SM with local gauged B and
L(BLMSSM) is more favorite\cite{ref9,ref10}. Since the new quarks
are vector-like with respect to the strong, weak and electromagnetic interactions to cancel anomalies, one obtains that their masses can be above 500 GeV without assuming large
couplings to the Higgs doublets in this model. Therefore, there are no Landau poles for the
Yukawa couplings here.

In BLMSSM, B and L are spontaneously broken near the weak scale, the proton decay is forbidden, and the three neutrinos get mass from the extended seesaw mechanism at tree level\cite{ref201,ref202,ref9,ref10}. Therefore, the desert between the grand unified scale and the electroweak scale is not necessary, which is the main motivation for the BLMSSM.

The CMS\cite{ref1001} and ATLAS\cite{ref1002} experiments of LHC have studied many possible signals of the MSSM, and set very strong bounds on the gluino and squarks masses with R-parity conservation. However, in the BLMSSM, the predictions and bounds for the collider experiments should be changed\cite{ref9,ref10,ref1003}. In addition, the lepton number violation could be detected at the LHC from the decays of right handed neutrinos\cite{ref201,ref202,ref1004}, and we could also look for the baryon number violation in the decays of squarks and gauginos\cite{ref1005}.
Since there are some exotic fields, and exist couplings between exotic quark fields and SM quarks in the superpotential, so it will cause flavor changing processes, and the BRs for FCNC top decays can be orders of magnitude larger.

In this paper  we analyze  the corrections  to the top-quark decay
$t\rightarrow ch$  in BLMSSM. This paper is composed of the sections
as follows. In section II, we present the main ingredients of the
BLMSSM. In section III, we present the theoretical calculation on
the $t\rightarrow ch$ processes. Section IV is devoted to the
numerical analysis. Our conclusions are
summarized in Section V.

\section{A supersymmtric extension of the SM where B and L are local gauge symmetries  \label{sec2}}
\indent\indent

The local gauge B and L is base on the gauge group:
$SU(3)_{_C}\otimes SU(2)_{_L}\otimes U(1)_{_Y}\otimes
U(1)_{_B}\otimes U(1)_{_L}$. In BLMSSM, to cancel the $B$ and $L$
anomalies, the exotic superfields should include the new quarks
$\hat{Q}_{_4}$, $\hat{U}_{_4}^c$, $\hat{D}_{_4}^c$,
$\hat{Q}_{_5}^c$, $\hat{U}_{_5}$, $\hat{D}_{_5}$, and the new
leptons $\hat{L}_{_4}$, $\hat{E}_{_4}^c$, $\hat{N}_{_4}^c$,
$\hat{L}_{_5}^c$, $\hat{E}_{_5}$, $\hat{N}_{_5}$. In addition, the
new Higgs chiral superfields $\hat{\Phi}_{_B}$ and
$\hat{\varphi}_{_B}$ acquire nonzero vacuum expectation values
(VEVs) to break baryon number spontaneously, the superfields
$\hat{\Phi}_{_L}$ and $\hat{\varphi}_{_L}$ acquire nonzero VEVs to
break lepton number spontaneously. The model also introduce the
superfields $\hat{X}$, $\hat{X}^\prime$ to avoid stability for the
exotic quarks. Actually, the lightest superfields can be a candidate
for dark matter . The properties  of these superfields in  BLMSSM
are summarized in Table I.

\begin{table}[!h]
\tabcolsep 0pt \caption{ The properties  of superfields in  BLMSSM} \vspace*{-12pt}
\begin{center}
\def\temptablewidth{0.6\textwidth}
{\rule{\temptablewidth}{1pt}}
\begin{tabular*}{\temptablewidth}{@{\extracolsep{\fill}}ccccccc}
superfield   & $SU(3)$ &  $SU(2)$ &  $U(1)_Y$ &$U(1)_B$ &$U(1)_L$ \\   \hline
     $\hat{Q}_{_4}$     &3           & 2  & 1/6  & $B_4$  &0  \\
       $\hat{U}_{_4}^c$ & $\bar{3}$  & 1  & -2/3 & $-B_4$ & 0 \\
      $\hat{D}_{_4}^c$  & $\bar{3}$  & 1  & 1/3  & $-B_4$ & 0  \\
        $\hat{Q}_{_5}^c$ & $\bar{3}$  & 2  & -1/6 & $-(1+B_4)$ & 0 \\
          $\hat{U}_{_5}$ & $3$  & 1  & 2/3 & $1+B_4$ & 0 \\
            $\hat{D}_{_5}$ & $3$  & 1  & -1/3 & $1+B_4$ & 0 \\
             $\hat{L}_{_4}$     &1          & 2  & -1/2  & 0  &$L_4$  \\
       $\hat{E}_{_4}^c$ & 1  & 1  & 1 & 0 & $-L_4$ \\
      $\hat{N}_{_4}^c$  & 1  & 1  & 0  & 0 & $-L_4$  \\
        $\hat{L}_{_5}^c$ & 1  & 2  & 1/2 & 0 & $-(3+L_4)$ \\
          $\hat{E}_{_5}$ & 1  & 1  & -1 & 0 & $3+L_4$\\
            $\hat{N}_{_5}$ & 1  & 1  &0 & 0 & $3+L_4$\\
             $\hat{\Phi}_{_B}$ & 1  & 1  &0 & 1& 0\\
             $\hat{\varphi}_{_B}$ & 1  & 1  &0 & -1& 0\\
             $\hat{\Phi}_{_L}$ & 1  & 1  &0 & 0& -2\\
             $\hat{\varphi}_{_L}$ & 1  & 1  &0 & 0& 2\\
              $\hat{X}$ & 1  & 1  &0 & $2/3+B_4$ &0\\
               $\hat{X}^\prime$ & 1  & 1  &0 & $-(2/3+B_4)$ & 0
       \end{tabular*}
       {\rule{\temptablewidth}{1pt}}
       \end{center}
       \end{table}

In BLMSSM, the super potential is written as \cite{ref11,ref12}
\begin{eqnarray}
&&{\cal W}_{_{BLMSSM}}={\cal W}_{_{MSSM}}+{\cal W}_{_B}+{\cal W}_{_L}+{\cal W}_{_X}\;,
\label{superpotential1}
\end{eqnarray}
where ${\cal W}_{_{MSSM}}$ is superpotential of the MSSM, and the
concrete form of ${\cal W}_{_B}$, ${\cal W}_{_L}$ and ${\cal
W}_{_X}$ are

\begin{eqnarray}
&&{\cal W}_{_B}=\lambda_{_Q}\hat{Q}_{_4}\hat{Q}_{_5}^c\hat{\Phi}_{_B}+\lambda_{_U}\hat{U}_{_4}^c\hat{U}_{_5}
\hat{\varphi}_{_B}+\lambda_{_D}\hat{D}_{_4}^c\hat{D}_{_5}\hat{\varphi}_{_B}+\mu_{_B}\hat{\Phi}_{_B}\hat{\varphi}_{_B}
\nonumber\\
&&\hspace{1.2cm}
+Y_{_{u_4}}\hat{Q}_{_4}\hat{H}_{_u}\hat{U}_{_4}^c+Y_{_{d_4}}\hat{Q}_{_4}\hat{H}_{_d}\hat{D}_{_4}^c
+Y_{_{u_5}}\hat{Q}_{_5}^c\hat{H}_{_d}\hat{U}_{_5}+Y_{_{d_5}}\hat{Q}_{_5}^c\hat{H}_{_u}\hat{D}_{_5}\;,
\nonumber\\
%%%%%%%%%%%%%%%%%%%%%%%%%%%%%%%%%%%%%%%%%%%%%%%%%%%%%%%%%%%%%
&&{\cal W}_{_L}=Y_{_{e_4}}\hat{L}_{_4}\hat{H}_{_d}\hat{E}_{_4}^c+Y_{_{\nu_4}}\hat{L}_{_4}\hat{H}_{_u}\hat{\nu}_{_4}^c
+Y_{_{e_5}}\hat{L}_{_5}^c\hat{H}_{_u}\hat{E}_{_5}+Y_{_{\nu_5}}\hat{L}_{_5}^c\hat{H}_{_d}\hat{\nu}_{_5}
\nonumber\\
&&\hspace{1.2cm}
+Y_{_\nu}\hat{L}\hat{H}_{_u}\hat{\nu}^c+\lambda_{_{\nu^c}}\hat{\nu}^c\hat{\nu}^c\hat{\varphi}_{_L}
+\mu_{_L}\hat{\Phi}_{_L}\hat{\varphi}_{_L}\;,
\nonumber\\
%%%%%%%%%%%%%%%%%%%%%%%%%%%%%%%%%%%%%%%%%%%%%%%%%%%%%%%%%%%%%
&&{\cal W}_{_X}=\lambda_1\hat{Q}\hat{Q}_{_5}^c\hat{X}+\lambda_2\hat{U}^c\hat{U}_{_5}\hat{X}^\prime
+\lambda_3\hat{D}^c\hat{D}_{_5}\hat{X}^\prime+\mu_{_X}\hat{X}\hat{X}^\prime\;,
\label{superpotential-BL}
\end{eqnarray}
and we could see that since ${\cal W}_{_X}$ contains superfields X and $Q_5$ ($U_5$, $D_5$ and $X^\prime$) couple to all generations of SM quarks, so FCNC processes can be generated.

Correspondingly, the soft breaking terms ${\cal L}_{_{soft}}$ are
generally given as
\begin{eqnarray}
&&{\cal L}_{_{soft}}={\cal L}_{_{soft}}^{MSSM}-(m_{_{\tilde{\nu}^c}}^2)_{_{IJ}}\tilde{\nu}_I^{c*}\tilde{\nu}_J^c
-m_{_{\tilde{Q}_4}}^2\tilde{Q}_{_4}^\dagger\tilde{Q}_{_4}-m_{_{\tilde{U}_4}}^2\tilde{U}_{_4}^{c*}\tilde{U}_{_4}^c
-m_{_{\tilde{D}_4}}^2\tilde{D}_{_4}^{c*}\tilde{D}_{_4}^c
\nonumber\\
&&\hspace{1.3cm}
-m_{_{\tilde{Q}_5}}^2\tilde{Q}_{_5}^{c\dagger}\tilde{Q}_{_5}^c-m_{_{\tilde{U}_5}}^2\tilde{U}_{_5}^*\tilde{U}_{_5}
-m_{_{\tilde{D}_5}}^2\tilde{D}_{_5}^*\tilde{D}_{_5}-m_{_{\tilde{L}_4}}^2\tilde{L}_{_4}^\dagger\tilde{L}_{_4}
-m_{_{\tilde{\nu}_4}}^2\tilde{\nu}_{_4}^{c*}\tilde{\nu}_{_4}^c
\nonumber\\
&&\hspace{1.3cm}
-m_{_{\tilde{E}_4}}^2\tilde{e}_{_4}^{c*}\tilde{e}_{_4}^c-m_{_{\tilde{L}_5}}^2\tilde{L}_{_5}^{c\dagger}\tilde{L}_{_5}^c
-m_{_{\tilde{\nu}_5}}^2\tilde{\nu}_{_5}^*\tilde{\nu}_{_5}-m_{_{\tilde{E}_5}}^2\tilde{e}_{_5}^*\tilde{e}_{_5}
-m_{_{\Phi_{_B}}}^2\Phi_{_B}^*\Phi_{_B}
\nonumber\\
&&\hspace{1.3cm}
-m_{_{\varphi_{_B}}}^2\varphi_{_B}^*\varphi_{_B}-m_{_{\Phi_{_L}}}^2\Phi_{_L}^*\Phi_{_L}
-m_{_{\varphi_{_L}}}^2\varphi_{_L}^*\varphi_{_L}-\Big(m_{_B}\lambda_{_B}\lambda_{_B}
+m_{_L}\lambda_{_L}\lambda_{_L}+h.c.\Big)
\nonumber\\
&&\hspace{1.3cm}
+\Big\{A_{_{u_4}}Y_{_{u_4}}\tilde{Q}_{_4}H_{_u}\tilde{U}_{_4}^c+A_{_{d_4}}Y_{_{d_4}}\tilde{Q}_{_4}H_{_d}\tilde{D}_{_4}^c
+A_{_{u_5}}Y_{_{u_5}}\tilde{Q}_{_5}^cH_{_d}\tilde{U}_{_5}+A_{_{d_5}}Y_{_{d_5}}\tilde{Q}_{_5}^cH_{_u}\tilde{D}_{_5}
\nonumber\\
&&\hspace{1.3cm}
+A_{_{BQ}}\lambda_{_Q}\tilde{Q}_{_4}\tilde{Q}_{_5}^c\Phi_{_B}+A_{_{BU}}\lambda_{_U}\tilde{U}_{_4}^c\tilde{U}_{_5}\varphi_{_B}
+A_{_{BD}}\lambda_{_D}\tilde{D}_{_4}^c\tilde{D}_{_5}\varphi_{_B}+B_{_B}\mu_{_B}\Phi_{_B}\varphi_{_B}
+h.c.\Big\}
\nonumber\\
&&\hspace{1.3cm}
+\Big\{A_{_{e_4}}Y_{_{e_4}}\tilde{L}_{_4}H_{_d}\tilde{E}_{_4}^c+A_{_{\nu_4}}Y_{_{\nu_4}}\tilde{L}_{_4}H_{_u}\tilde{\nu}_{_4}^c
+A_{_{e_5}}Y_{_{e_5}}\tilde{L}_{_5}^cH_{_u}\tilde{E}_{_5}+A_{_{\nu_5}}Y_{_{\nu_5}}\tilde{L}_{_5}^cH_{_d}\tilde{\nu}_{_5}
\nonumber\\
&&\hspace{1.3cm}
+A_{_\nu}Y_{_\nu}\tilde{L}H_{_u}\tilde{\nu}^c+A_{_{\nu^c}}\lambda_{_{\nu^c}}\tilde{\nu}^c\tilde{\nu}^c\varphi_{_L}
+B_{_L}\mu_{_L}\Phi_{_L}\varphi_{_L}+h.c.\Big\}
\nonumber\\
&&\hspace{1.3cm}
+\Big\{A_1\lambda_1\tilde{Q}\tilde{Q}_{_5}^cX+A_2\lambda_2\tilde{U}^c\tilde{U}_{_5}X^\prime
+A_3\lambda_3\tilde{D}^c\tilde{D}_{_5}X^\prime+B_{_X}\mu_{_X}XX^\prime+h.c.\Big\}\;,
\label{soft-breaking}
\end{eqnarray}
with ${\cal L}_{_{soft}}^{MSSM}$ representing the soft breaking
terms of the MSSM, and $\lambda_B,\;\lambda_L$ are gauginos of
$U(1)_{_B}$ and $U(1)_{_L}$, respectively.

To  break the local gauge symmetry $SU(2)_{_L}\otimes U(1)_{_Y}\otimes
U(1)_{_B}\otimes U(1)_{_L}$ down  to the electromagnetic
symmetry $U(1)_{_e}$, the $SU(2)_L$ doublets $H_{_u},\;H_{_d}$ and
the $SU(2)_L$ singlets $\Phi_{_B},\;\varphi_{_B},\;\Phi_{_L},\;
\varphi_{_L}$ should obtain the nonzero VEVs
$\upsilon_{_u},\;\upsilon_{_d},\;\upsilon_{_{B}},\;\overline{\upsilon}_{_{B}}$,
and $\upsilon_{_L},\;\overline{\upsilon}_{_L}$ respectively.

\begin{eqnarray}
&&H_{_u}=\left(\begin{array}{c}H_{_u}^+\\{1\over\sqrt{2}}\Big(\upsilon_{_u}+H_{_u}^0+iP_{_u}^0\Big)\end{array}\right)\;,
\nonumber\\
&&H_{_d}=\left(\begin{array}{c}{1\over\sqrt{2}}\Big(\upsilon_{_d}+H_{_d}^0+iP_{_d}^0\Big)\\H_{_d}^-\end{array}\right)\;,
\nonumber\\
&&\Phi_{_B}={1\over\sqrt{2}}\Big(\upsilon_{_B}+\Phi_{_B}^0+iP_{_B}^0\Big)\;,
\nonumber\\
&&\varphi_{_B}={1\over\sqrt{2}}\Big(\overline{\upsilon}_{_B}+\varphi_{_B}^0+i\overline{P}_{_B}^0\Big)\;,
\nonumber\\
&&\Phi_{_L}={1\over\sqrt{2}}\Big(\upsilon_{_L}+\Phi_{_L}^0+iP_{_L}^0\Big)\;,
\nonumber\\
&&\varphi_{_L}={1\over\sqrt{2}}\Big(\overline{\upsilon}_{_L}+\varphi_{_L}^0+i\overline{P}_{_L}^0\Big)\;,
\label{VEVs}
\end{eqnarray}

The mass matrixes of Higgs, exotic quarks and exotic scalar quarks
are obtained in our previous work\cite{ref11}, and we list some
useful results.

In four-component Dirac spinors, the mass matrix for exotic charged
2/3 quarks is
\begin{eqnarray}
&& - L_{{t'' }}^{mass} = \left( {\begin{array}{*{20}{l}} {\bar
t''_{4R},}&{\bar t''_{5R}}
\end{array}} \right)\left( {\begin{array}{*{20}{l}}
{ \frac{1}{{\sqrt 2 }}{Y_{{u_4}}}{\upsilon _u},}&{-\frac{1}{{\sqrt 2 }}{\lambda _Q}{\upsilon _B}}\\
{- \frac{1}{{\sqrt
2 }}{\lambda _u}{{\bar \upsilon }_B},}
&{  \frac{1}{{\sqrt 2 }}{Y_{{u_5}}}{\upsilon _d}}
\end{array}} \right)\left( {\begin{array}{*{20}{l}}
{t''_{4L}}\\
{t''_{5L}}
\end{array}} \right) + h.c.\label{VEVs}
\end{eqnarray}
and it could be diagonalized by the the unitary transformations
\begin{eqnarray}
&&\left( {\begin{array}{*{20}{l}}
{t'_{4L}}\\
{t'_{5L}}
\end{array}} \right) = U_{{t^\prime }}^\dag \cdot\left( {\begin{array}{*{20}{l}}
{{{t''}_{4L}}}\\
{{{t''}_{5L}}}
\end{array}} \right)\;,\;\;\left( {\begin{array}{*{20}{l}}
{t'_{4R}}\\
{t'_{5R}}
\end{array}} \right) = W_{{t^\prime }}^\dag \cdot\left( {\begin{array}{*{20}{l}}
{{{t''}_{4R}}}\\
{{{t''}_{5R}}}
\end{array}} \right)\;,\label{VEVs}
\end{eqnarray}
then we get
\begin{eqnarray}
&&W_{{t^\prime }}^\dag \cdot\left( {\begin{array}{*{20}{l}}
{ \frac{1}{{\sqrt 2 }}{Y_{{u_4}}}{\upsilon _u},}&{-\frac{1}{{\sqrt 2 }}{\lambda _Q}{\upsilon _B}}\\
{- \frac{1}{{\sqrt
2 }}{\lambda _u}{{\bar \upsilon }_B},}
&{  \frac{1}{{\sqrt 2 }}{Y_{{u_5}}}{\upsilon _d}}
\end{array}} \right)\cdot{U_{{t^\prime }}} = diag({m_{{t_4}}},\;{m_{{t_5}}})\label{VEVs}
\end{eqnarray}

Similarly, The concrete expressions for $4\times4$ mass squared
matrice $M_{\tilde t'}^2$ of exotic charged 2/3 scalar quarks
${\tilde t''^T} = (\tilde Q_4^1,\;\tilde U_4^{c*},\;\tilde
Q_5^{2c*},\;{\tilde U_5})$ are given in appendix B of
Ref\cite{ref11}, and it could be diagonalized by the the unitary
transformation
\begin{eqnarray}
&&{\tilde t_i}^{\prime \prime } = Z_{\tilde t'}^{ij}{\tilde
t'_j},\label{Higgs-EQ-1/3}
\end{eqnarray}

Using the scalar potential and the soft breaking terms, the mass
squared matrix for $X,X'$ could be written as

\begin{eqnarray}
&&-L_X^{mass}=\left( {\begin{array}{*{20}{c}} {X^*}&{X'}
\end{array}} \right)\left( {\begin{array}{*{20}{c}}
{\mu _X^2+S_X}&{-{B_X}{\mu _X}}\\
{-{B_X}{\mu _X}}&{\mu _X^2-S_X}
\end{array}} \right)\left( {\begin{array}{*{20}{c}}
{X}\\
{X'}
\end{array}} \right)\label{Higgs-EQ-1/3}
\end{eqnarray}
with $S_X=\frac{g_B^2}{2}(\frac{2}{3}+B_4)(v_B^2-\overline{v}_B^2)$.
And it could be diagonalized by the unitary transformation $Z_X$
\begin{eqnarray}
&&Z_X^\dag \left( {\begin{array}{*{20}{c}}
{\mu _X^2+S_X}&{-{B_X}{\mu _X}}\\
{-{B_X}{\mu _X}}&{\mu _X^2-S_X}
\end{array}} \right){Z_X} = diag(m_{{X_1}}^2,\;m_{{X_2}}^2)\;.\label{Higgs-EQ-1/3}
\end{eqnarray}

In addition, the four-component Dirac spinor $\widetilde{X}$ is defined as  $\widetilde{X}=(\psi_X, \bar{\psi}_{X'})^T$, with the mass term $\mu_X\widetilde{X}\overline{\widetilde{X}}$.

The flavor conservative couplings between the lightest neutral
Higgs and charged $2/3$ exotic quarks are
\begin{eqnarray}
&&{\cal L}_{{Ht^\prime
t^\prime}}={1\over\sqrt{2}}\sum\limits_{i,j=1}^2\Big\{
\Big[Y_{{u_4}}(W_{t}^\dagger)_{{i2}}(U_{t})_{{1j}}\cos\alpha
+Y_{{u_5}}(W_{t}^\dagger)_{{i1}}(U_{t})_{{2j}}\sin\alpha\Big]
h^0\overline{t'}_{{i}}P_{L}t'_{{j}}
\nonumber\\
&&\hspace{1.8cm}
+\Big[Y_{{u_4}}(U_{t}^\dagger)_{{i1}}(W_{t})_{{2j}}\cos\alpha
+Y_{{u_5}}(U_{t}^\dagger)_{{i2}}(W_{t})_{{1j}}\sin\alpha\Big]
h^0\overline{t'}_{{i}}P_{R}t'_{{j}}\label{Higgs-EQ-2/3}
\end{eqnarray}
with $\alpha$ is defined as
 \begin{eqnarray}
&&\left(\begin{array}{l}H^0\\h^0\end{array}\right)=\left(\begin{array}{cc}\cos\alpha&\sin\alpha\\
-\sin\alpha&\cos\alpha\end{array}\right)\left(\begin{array}{l}H_{_d}^0\\H_{_u}^0\end{array}\right)\;,
\label{charged-Higgs}
\end{eqnarray}
And the couplings  between the lightest neutral
 Higgs and exotic scalar quarks are

\begin{eqnarray}
&&{\cal L}_{{H\tilde {t'}_i^*\tilde
{t'}}}=\sum\limits_{i,j}^4\Big[\xi_{{uij}}^S\cos\alpha
-\xi_{{dij}}^S\sin\alpha\Big]h^0\tilde {t'}_i^*\tilde {t'}_j
\label{Higgs-EQ-2/3}
\end{eqnarray}
with $\xi_{{uij}}^S$ and $\xi_{{dij}}^S$ are defined in  appendix C
of Ref\cite{ref11}.

 In mass basis, we obtain the couplings of
quark-exotic quark and the  $X$:
\begin{eqnarray}
&&- {\lambda _1}{({W_{t'}})_{i2}}{({Z_X})_{1j}}{X_j}{\kern 1pt}
{\bar t'_i}{P_L}u - {\lambda _2}{(U_{t'}^\dag
)_{2i}}{({Z_X})_{2j}}{X_j}{\kern 1pt} {\kern 1pt} \bar u{P_L}{t'_i}
+ h.c.\label{Higgs-EQ-1/3}
\end{eqnarray}
and the couplings between up type quark and the superpartners
$\widetilde{t}', \widetilde{X}$ are
\begin{eqnarray}
&&- {\lambda _1}{(Z_{\tilde t'}^\dag )_{i3}}{\tilde t'_i}{\kern 1pt}
\bar u{P_L}\tilde X - {\lambda _2}{(Z_{\tilde t'}^\dag
)_{i4}}{\tilde t'_i}\bar{\tilde{X}}{P_L}u + h.c.\label{Higgs-EQ-1/3}
\end{eqnarray}

\section{The theoretical calculation on the  $t\rightarrow ch$ process  \label{sec2}}
\indent\indent

In this section, we present one-loop radiative corrections to the
rare decay $t\rightarrow ch$ in BLMSSM. For this process, it is
convenient to define an effective interaction vertex \cite{ref6}:
\begin{eqnarray}
&&- iT =  - ig\bar c(p)\left( {{F_L}{P_L} + {F_R}{P_R}} \right)t(p')
\label{Higgs-EQ-1/3}
\end{eqnarray}
where $p'$ is the momentum of the  initial top quark, $p$ is the
momentum of the final state charm quark, and form factors $F_L$, $F_R$ are  follow from explicit calculation of vertices and mixed self-energies.

 The relevant one-loop vertex diagrams
of BLMSSM are drawn in Fig.1.

%%%%%%%%%%%%%%%%%%%%%%%%%%%%%%%%%%%%%%%%%%%%%%%%%%%%%%%%%%%%%%%%%%%%%%%%%%%%%%%%%%%%5
\begin{figure}\small
  \centering
   \includegraphics[width=6in]{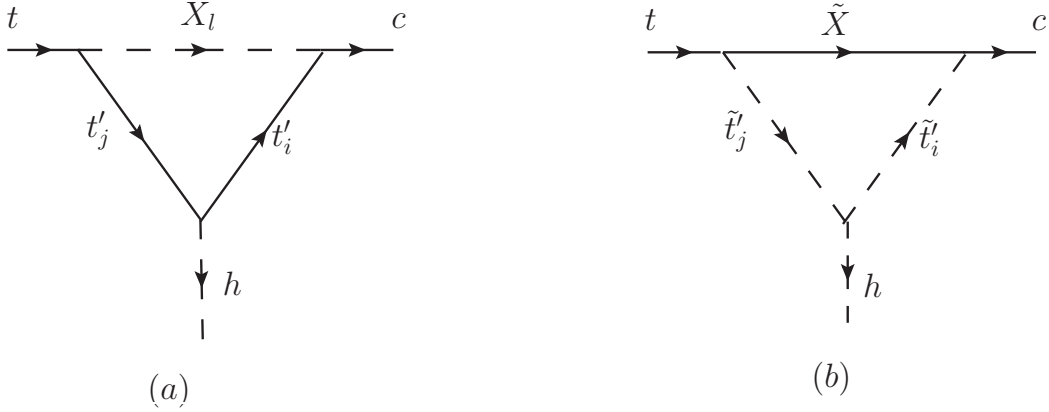}
     \caption{The vertex diagrams contributing to the
$t\rightarrow ch$ decay in BLMSSM. }
    \label{fige3c}
\end{figure}
%%%%%%%%%%%%%%%%%%%%%%%%%%%%%%%%%%%%%%%%%%%%%%%%%%%%%%%%%%%%%%%%%%%%%%%%%%%%%%%%%%%%%%%%
We could see that the FCNC transitions of new physics are mediated
by the exotic up type quark  $t'$, the neutral scalar particle $X_i$
and there superpartners $\widetilde{t}',  \widetilde{X}$. And the contribution to the form factors could be obtained by direct calculation.

In the equations below, $m_{t'}$, $m_X$, $m_{\widetilde{t}'}$,
$m_{\widetilde{X}}$  denotes the mass of the exotic quarks $t'$, the
mass of scalar particle $X_i$, and the mass of there superpartners
$\widetilde{t}',  \widetilde{X}$  respectively. $B_{i},{C}_{ij}$ are
the coefficients of the Lorentz-covariant tensors in the standard
scalar Passarino-Veltman integrals(Eq.(4.7) in Ref. \cite{ref13}),
and it could be calculated by using 'LoopTools'.

In Fig.1(a), when one-loop diagrams are composed by the neutral
scalar particles $X_i$, and charged $2/3$ new quarks $t' $, the
contribution to the form factors $F_{L}^{a}$ and $F_{R}^{a}$ are
formulated as
\begin{eqnarray}
&&F_L^a =\frac{i}{{16{\pi ^2}}}\sum_{i,j,\ l}( - {a_1}{m_c}({b_1}{h_2}{m_t}{C_2} + {b_2}{h_1}{m_{{{t'}_i}}}({C_0} + {C_1} + 2{C_2}) + 3{b_2}{h_2}{m_{{{t'}_j}}}{C_2}))\nonumber\\
&&\hspace{1.3cm}+ {a_2}{b_2}({h_1}{B_0} + ({h_1}m_{{{t'}_i}}^2 + {h_2}{m_{{{t'}_i}}}{m_{{{t'}_j}}}){C_0})\nonumber\\
&&\hspace{1.3cm} + {a_2}{b_1}{m_t}({h_2}{m_{{{t'}_i}}}({C_0} + {C_1} + {C_2}) + {h_1}{m_{{{t'}_j}}}({C_1} + {C_2}))+ {a_2}{b_2}{h_1}{m_c}^2{C_2}\nonumber\\
&&F_R^a =\frac{i}{{16{\pi ^2}}}\sum_{i,j,\ l}( -
{a_2}{m_c}({b_1}{h_2}{m_{{{t'}_i}}}({C_0} + {C_1} + 2{C_2}) +
{b_1}{h_1}{m_{{{t'}_j}}}({C_1} + 2{C_2}) +
{b_2}{h_1}{m_t}{C_2}))\nonumber\\
&&\hspace{1.3cm}+ {a_1}{b_1}({h_2}{B_0} + ({h_1}{m_{{{t'}_i}}}{m_{{{t'}_j}}} + {h_2}m_{{{t'}_i}}^2){C_0}) \nonumber\\
&&\hspace{1.3cm}+ {a_1}{b_2}{m_t}({h_1}{m_{{{t'}_i}}}({C_0} + {C_1} + {C_2}) + {h_2}{m_{{{t'}_j}}}({C_1} + {C_2}))+ {a_1}{b_1}{h_2}{m_c}^2{C_2}\label{Higgs-EQ-1/3}
\end{eqnarray}
with the Passarino-Veltman integrals
\begin{eqnarray}
&&{B_0} = {B_0}({p^2},m_{{{t'}_j}}^2,m{_{X_l}^2})
\nonumber\\
&&{{\rm{C}}_0}{\rm{ = }}{{\rm{C}}_0}\left( {{p^2},{{(2p -
p')}^2},{{(p - p')}^2},m_{{{t'}_j}}^2,m_{{X_l}}^2,m_{{{t'}_i}}^2}
\right)\nonumber\\
&&{{\rm{C}}_{1,2}}{\rm{ = }}{{\rm{C}}_{1,2}}\left( {{{(p -
p')}^2},{{(2p -
p')}^2},{p^2},m_{{{t'}_i}}^2,m_{{{t'}_j}}^2,m_{{X_l}}^2} \right)
\label{Higgs-EQ-1/3}
\end{eqnarray}
and the relevant coefficients are
\begin{eqnarray}
&&{a_1} = \lambda _1^*{(W_{t'}^\dag )_{2i}}{(Z_X^\dag )_{l1}},{\kern
1pt} {\kern 1pt} {\kern 1pt} {\kern 1pt} {\kern 1pt} {\kern 1pt}
{\kern 1pt} {\kern 1pt} {\kern 1pt} {\kern 1pt} {\kern 1pt} {\kern
1pt} {\kern 1pt} {\kern 1pt} {\kern 1pt} {\kern 1pt} {\kern 1pt}
{\kern 1pt} {\kern 1pt} {\kern 1pt} {\kern 1pt} {\kern 1pt} {\kern
1pt} {\kern 1pt} {\kern 1pt} {\kern 1pt} {\kern 1pt} {a_2} =
{\lambda _2}{(U_{t'}^\dag )_{2i}}{({Z_X})_{2l}},\nonumber\\
 &&{b_1}
= \lambda _2^*{({U_{t'}})_{j2}}{(Z_X^\dag )_{l2}},{\kern 1pt} {\kern
1pt} {\kern 1pt} {\kern 1pt} {\kern 1pt} {\kern 1pt} {\kern 1pt}
{\kern 1pt} {\kern 1pt} {\kern 1pt} {\kern 1pt} {\kern 1pt} {\kern
1pt} {\kern 1pt} {\kern 1pt} {\kern 1pt} {\kern 1pt} {\kern 1pt}
{\kern 1pt} {\kern 1pt} {\kern 1pt} {\kern 1pt} {\kern 1pt} {\kern
1pt} {\kern 1pt} {\kern 1pt} {\kern 1pt} {\kern 1pt} {b_2} =
{\lambda _1}{({W_{t'}})_{j2}}{({Z_X})_{1l}},\nonumber\\
 &&{h_1} =
{Y_{{u_4}}}{(U_{t'}^\dag )_{i1}}{({W_{t'}})_{2j}}\cos \alpha  +
{Y_{{u_5}}}{(U_{t'}^\dag )_{i2}}{({W_{t'}})_{1j}}\sin \alpha
,\nonumber\\
 &&{h_2} = {Y_{{u_4}}}{(W_{t'}^\dag
)_{i2}}{({U_{t'}})_{1j}}\cos \alpha + {Y_{{u_5}}}{(W_{t'}^\dag
)_{i1}}{({U_{t'}})_{2j}}\sin \alpha , \label{Higgs-EQ-1/3}
\end{eqnarray}

In Fig.1(b), when one-loop diagrams are composed by the
superpartners $\widetilde{t}'$ and $\widetilde{X}$, $F_{L}^{b}$ and
$F_{R}^{b}$ are formulated as

\begin{eqnarray}
&&F_L^b =\frac{i}{{16{\pi ^2}}}\sum_{i,j} ({a_4}{b_4}{m_{\tilde X}}{C_0} -
{a_3}{b_4}{m_c}{C_1} - {a_4}{b_3}{m_t}{C_2})(\cos {\alpha }{\xi
_u} - \sin {\alpha }{\xi _d})\nonumber\\
 &&F_R^b = \frac{i}{{16{\pi ^2}}}\sum_{i,j} ({a_4}{b_4}{m_{\tilde X}}{C_0} - {a_3}{b_4}{m_c}{C_1} - {a_4}{b_3}{m_t}{C_2})(\cos {\alpha }{\xi _u} - \sin {\alpha }{\xi _d})\label{Higgs-EQ-1/3}
\end{eqnarray}
with
\begin{eqnarray}
&&{{\rm{C}}_0}{\rm{ = }}{{\rm{C}}_0}\left( {{p^2},{{p'}^2},{{(p -
p')}^2},m_{{{\tilde t'}_i}}^2,m_{\tilde X}^2,m_{{{\tilde t'}_j}}^2}
\right)\nonumber\\
 &&{{\rm{C}}_{1,2}}{\rm{ = }}{{\rm{C}}_{1,2}}\left( {{p^2},{{(p -
p')}^2},{{p'}^2},m_{\tilde X}^2,m_{{{\tilde t'}_i}}^2,m_{{{\tilde
t'}_j}}^2} \right)\label{Higgs-EQ-1/3}
\end{eqnarray}
and the relevant coefficients are
\begin{eqnarray}
&&
{a_3} = \lambda _2^*{(Z_{\tilde t'}^\dag )_{i4}},{\kern 1pt} {\kern 1pt} {\kern 1pt} {\kern 1pt} {\kern 1pt} {\kern 1pt} {\kern 1pt} {\kern 1pt} {\kern 1pt} {\kern 1pt} {\kern 1pt} {\kern 1pt} {\kern 1pt} {\kern 1pt} {\kern 1pt} {\kern 1pt} {\kern 1pt} {\kern 1pt} {\kern 1pt} {\kern 1pt} {\kern 1pt} {\kern 1pt} {\kern 1pt} {\kern 1pt} {\kern 1pt} {\kern 1pt} {\kern 1pt} {\kern 1pt} {\kern 1pt} {\kern 1pt} {\kern 1pt} {\kern 1pt} {\kern 1pt} {\kern 1pt} {\kern 1pt} {\kern 1pt} {\kern 1pt} {a_4} = {\lambda _1}{(Z_{\tilde t'}^\dag )_{i3}},\nonumber\\
 &&
{b_3} = \lambda _1^*{({Z_{\tilde t'}})_{3j}},{\kern 1pt} {\kern 1pt}
{\kern 1pt} {\kern 1pt} {\kern 1pt} {\kern 1pt} {\kern 1pt} {\kern
1pt} {\kern 1pt} {\kern 1pt} {\kern 1pt} {\kern 1pt} {\kern 1pt}
{\kern 1pt} {\kern 1pt} {\kern 1pt} {\kern 1pt} {\kern 1pt} {\kern
1pt} {\kern 1pt} {\kern 1pt} {\kern 1pt} {\kern 1pt} {\kern 1pt}
{\kern 1pt} {\kern 1pt} {\kern 1pt} {\kern 1pt} {\kern 1pt} {\kern
1pt} {\kern 1pt} {\kern 1pt} {\kern 1pt} {\kern 1pt} {\kern 1pt}
{\kern 1pt} {\kern 1pt} {b_4} = {\lambda _2}{({Z_{\tilde
t'}})_{4j}}, \label{Higgs-EQ-1/3}
\end{eqnarray}

In Fig.2 we present the relevant self-energy diagrams of the rare
decay $t\rightarrow ch$ in BLMSSM.

%%%%%%%%%%%%%%%%%%%%%%%%%%%%%%%%%%%%%%%%%%%%%%%%%%%%%%%%%%%%%%%%%%%%%%%%%%%%%%%%%%%%5
\begin{figure}\small
  \centering
   \includegraphics[width=6in]{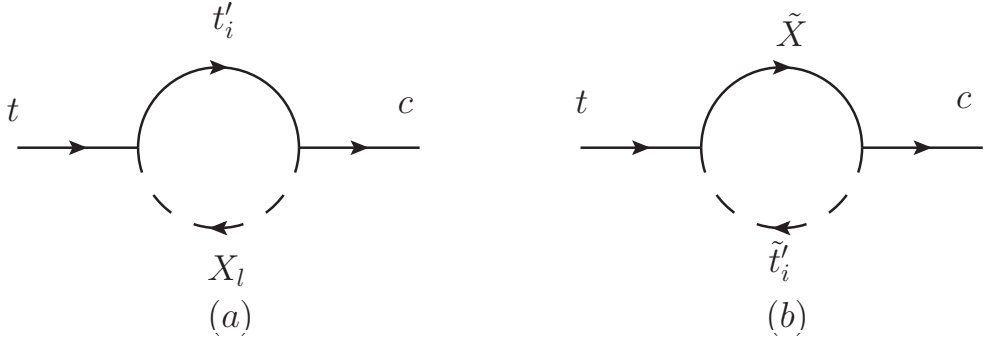}
     \caption{The self-energy diagrams contributing to the
$t\rightarrow ch$ decay in BLMSSM. }
    \label{fige3c}
\end{figure}
%%%%%%%%%%%%%%%%%%%%%%%%%%%%%%%%%%%%%%%%%%%%%%%%%%%%%%%%%%%%%%%%%%%%%%%%%%%%%%%%%%%%%%%%

As in Ref.\cite{ref6}, it is convenient to define the following
structure:
\begin{eqnarray}
&&{\Sigma _{tc}}(k) \equiv \not {k}{\Sigma _L}({k^2}){P_L} +\not k
{\Sigma _R}({k^2}){P_R} + {m_t}({\Sigma _{Ls}}({k^2}){P_L} + {\Sigma
_{Rs}}({k^2}){P_R}).\label{Higgs-EQ-1/3}
\end{eqnarray}
Here $m_t$ factor is inserted there only to preserve the same
dimensionality for the different $\Sigma$\cite{ref6}. And the
effective interaction vertex of the mixed self-energy diagrams could
be taken on the following general form in terms of the various
$\Sigma$.

\begin{eqnarray}
&&- i{T_{Sc}} = \frac{{ - ig{m_t}}}{{2{m_W}\sin \beta }}\frac{1}{{m_c^2 - m_t^2}}\bar c(p)\left\{ {} \right.\nonumber\\
 &&
{\kern 1pt} {\kern 1pt} {\kern 1pt} {\kern 1pt} {\kern 1pt} {\kern 1pt} {\kern 1pt} {\kern 1pt} {\kern 1pt} {\kern 1pt} {\kern 1pt} {\kern 1pt} {\kern 1pt} {\kern 1pt} {\kern 1pt} {\kern 1pt} {\kern 1pt} {\kern 1pt} {\kern 1pt} {\kern 1pt} {\kern 1pt} {\kern 1pt} {\kern 1pt} {\kern 1pt} {\kern 1pt} {\kern 1pt} {\kern 1pt} {\kern 1pt} {\kern 1pt} {\kern 1pt} {\kern 1pt} {\kern 1pt} {\kern 1pt} {\kern 1pt} {\kern 1pt} ({P_L}\cos \alpha [m_c^2{\Sigma _R}(m_c^2) + {m_c}{m_t}({\Sigma _{Rs}}(m_c^2) + {\Sigma _L}(m_c^2)) + m_t^2{\Sigma _{Ls}}(m_c^2)]\nonumber\\
 &&
\left. {{\kern 1pt} {\kern 1pt} {\kern 1pt} {\kern 1pt} {\kern 1pt} {\kern 1pt} {\kern 1pt} {\kern 1pt} {\kern 1pt} {\kern 1pt} {\kern 1pt} {\kern 1pt} {\kern 1pt} {\kern 1pt} {\kern 1pt} {\kern 1pt} {\kern 1pt} {\kern 1pt} {\kern 1pt} {\kern 1pt} {\kern 1pt} {\kern 1pt} {\kern 1pt} {\kern 1pt} {\kern 1pt} {\kern 1pt} {\kern 1pt} {\kern 1pt} {\kern 1pt} {\kern 1pt} {\kern 1pt} {\kern 1pt} {\kern 1pt}  + {P_R}\cos \alpha [L \leftrightarrow R]} \right\}t(p')\nonumber\\
 &&
 - i{T_{St}} = \frac{{ - ig{m_c}}}{{2{m_W}\sin \beta }}\frac{{{m_t}}}{{m_c^2 - m_t^2}}\bar c(p)\left\{ {} \right.\nonumber\\
 &&
{\kern 1pt} {\kern 1pt} {\kern 1pt} {\kern 1pt} {\kern 1pt} {\kern 1pt} {\kern 1pt} {\kern 1pt} {\kern 1pt} {\kern 1pt} {\kern 1pt} {\kern 1pt} {\kern 1pt} {\kern 1pt} {\kern 1pt} {\kern 1pt} {\kern 1pt} {\kern 1pt} {\kern 1pt} {\kern 1pt} {\kern 1pt} {\kern 1pt} {\kern 1pt} {\kern 1pt} {\kern 1pt} {\kern 1pt} {\kern 1pt} {\kern 1pt} {\kern 1pt} {\kern 1pt} {\kern 1pt} {\kern 1pt} {\kern 1pt} {\kern 1pt} ({P_L}\cos \alpha [{m_t}({\Sigma _L}(m_t^2) + {\Sigma _{Rs}}(m_t^2)) + {m_c}({\Sigma _R}(m_t^2) + {\Sigma _{Ls}}(m_t^2))]\nonumber\\
 &&
{\kern 1pt} {\kern 1pt} {\kern 1pt} {\kern 1pt} {\kern 1pt} {\kern
1pt} {\kern 1pt} {\kern 1pt} {\kern 1pt} {\kern 1pt} {\kern 1pt}
{\kern 1pt} {\kern 1pt} {\kern 1pt} {\kern 1pt} {\kern 1pt} {\kern
1pt} {\kern 1pt} {\kern 1pt} {\kern 1pt} {\kern 1pt} {\kern 1pt}
{\kern 1pt} {\kern 1pt} {\kern 1pt} {\kern 1pt} {\kern 1pt} {\kern
1pt} {\kern 1pt} {\kern 1pt} {\kern 1pt} {\kern 1pt} {\kern 1pt}
{\kern 1pt} \left. { + {P_R}\cos \alpha [L \leftrightarrow R]}
\right\}t(p') \label{Higgs-EQ-1/3}
\end{eqnarray}
Comparing with Eq. 16, the corresponding contribution to the form
factors $F_{L}$ and $F_{R}$ is transparent.

Using the couplings above, we could get the $\Sigma$ of
self-energy diagrams in Fig.2(a) is

\begin{eqnarray}
&&{\Sigma _L}({k^2}) = \frac{i}{{16{\pi
^2}}}\sum_{i,\ l}{a_1}{b_2}({B_0}({k^2},m_{{X_l}}^2,m_{t'}^2) +
{B_1}({k^2},m_{{X_l}}^2,m_{t'}^2))\nonumber\\
 &&{\Sigma _R}({k^2}) = \frac{i}{{16{\pi ^2}}}\sum_{i,\ l}{a_2}{b_1}({B_0}({k^2},m_{{X_l}}^2,m_{t'}^2) + {B_1}({k^2},m_{{X_l}}^2,m_{t'}^2))\nonumber\\
 &&{m_t}{\Sigma _{Ls}}({k^2}) = \frac{i}{{16{\pi ^2}}}\sum_{i,\ l}{a_2}{b_2}{m_{t'}}{B_0}({k^2},m_{{X_l}}^2,m_{t'}^2)\nonumber\\
 &&{m_t}{\Sigma _{Rs}}({k^2}) = \frac{i}{{16{\pi ^2}}}\sum_{i,\ l}{a_1}{b_1}{m_{t'}}{B_0}({k^2},m_{{X_l}}^2,m_{t'}^2)
\label{Higgs-EQ-1/3}
\end{eqnarray}
with $B_{0,1}$ are  the two-point functions. Similarly, the  the
$\Sigma$ of self-energy diagrams in Fig.2(b) have the form:
\begin{eqnarray}
&&{\Sigma _L}({k^2}) = \frac{i}{{16{\pi
^2}}}\sum_{i}{a_3}{b_4}({B_0}({k^2},m_{\tilde t'}^2,m_{\tilde X}^2) +
{B_1}({k^2},m_{\tilde t'}^2,m_{\tilde X}^2))\nonumber\\
 &&
{\Sigma _R}({k^2}) = \frac{i}{{16{\pi ^2}}}\sum_{i}{a_4}{b_3}({B_0}({k^2},m_{\tilde t'}^2,m_{\tilde X}^2) + {B_1}({k^2},m_{\tilde t'}^2,m_{\tilde X}^2))\nonumber\\
 &&{m_t}{\Sigma _{Ls}}({k^2}) = \frac{i}{{16{\pi ^2}}}\sum_{i}{a_4}{b_4}{m_{\tilde t'}}{B_0}({k^2},m_{\tilde t'}^2,m_{\tilde X}^2)\nonumber\\
 &&{m_t}{\Sigma _{Rs}}({k^2}) = \frac{i}{{16{\pi ^2}}}\sum_{i}{a_3}{b_3}{m_{\tilde t'}}{B_0}({k^2},m_{\tilde t'}^2,m_{\tilde X}^2)\label{Higgs-EQ-1/3}
\end{eqnarray}
\section{Numerical analysis \label{sec3}}
\indent\indent

In general case, the partial widths of $t\rightarrow ch$ process
are\cite{ref6}
\begin{eqnarray}
&&
\Gamma (t \to ch) = \frac{{{g^2}}}{{32\pi m_t^3}}{\lambda ^{1/2}}(m_t^2,m_h^2,m_c^2)\nonumber\\
 &&
{\kern 1pt} {\kern 1pt} {\kern 1pt} {\kern 1pt} {\kern 1pt} {\kern
1pt} {\kern 1pt} {\kern 1pt} {\kern 1pt} {\kern 1pt} {\kern 1pt}
{\kern 1pt} {\kern 1pt} {\kern 1pt} {\kern 1pt} {\kern 1pt} {\kern
1pt} {\kern 1pt} {\kern 1pt} {\kern 1pt} {\kern 1pt} {\kern 1pt}
{\kern 1pt} {\kern 1pt} {\kern 1pt} {\kern 1pt} {\kern 1pt} {\kern
1pt} {\kern 1pt} {\kern 1pt} {\kern 1pt} {\kern 1pt} {\kern 1pt}
{\kern 1pt} {\kern 1pt} {\kern 1pt} {\kern 1pt} {\kern 1pt} {\kern
1pt} {\kern 1pt} {\kern 1pt} {\kern 1pt} {\kern 1pt} {\kern 1pt}
{\kern 1pt} {\kern 1pt} {\kern 1pt} {\kern 1pt} {\kern 1pt} {\kern
1pt} {\kern 1pt} {\kern 1pt} {\kern 1pt} {\kern 1pt} {\kern 1pt}
{\kern 1pt} {\kern 1pt} {\kern 1pt} {\kern 1pt} {\kern 1pt}  \times
\left[ {\left( {m_t^2 + m_c^2 - m_h^2} \right)\left( {{{\left|
{{F_L}} \right|}^2} + {{\left| {{F_R}} \right|}^2}} \right) +
2{m_t}{m_c}\left( {{F_L}F_R^* + F_L^*{F_R}} \right)} \right]
\label{Higgs-EQ-1/3}
\end{eqnarray}
with $\lambda ({x^2},{y^2},{z^2}) = ({x^2} - {(y + z)^2})({x^2} -
{(y - z)^2})$ is the usual K\"{a}llen function, and
\begin{eqnarray}
&&
{F_L} = F_L^{BLSSM} + F_L^{MSSM} + F_L^{SM}\nonumber\\
 &&
{F_R} = F_R^{BLSSM} + F_R^{MSSM} + F_R^{SM} \label{Higgs-EQ-1/3}
\end{eqnarray}
In our calculation,we will use the form factors of MSSM $F_{L,R}^{MSSM}$ mentioned in\cite{ref6}. And since the contributions of SM is too small, about $10^{-13}$\cite{ref5}, so we ignore the form factors of SM.

To compute the branching ratio, we take the SM charged-current
two-body decay $t\rightarrow bW$ to be the dominant $t$-quark decay
mode, which is $\Gamma(t\rightarrow bW^+)=1.466|V_{tb}|^2$ . We will
then approximate the branching ratio by

\begin{eqnarray}
&&Br(t \to ch) = \frac{{\Gamma (t \to ch)}}{{\Gamma (t \to b{W^ +
})}}\label{Higgs-EQ-1/3}
\end{eqnarray}

To reduce the number of free parameters in our numerical analysis,
the  parameters are adopted as Ref.\cite{ref11,ref12}. In this choice,  it is easy for the
$2\times2$ CP-even Higgs mass squared matrix to predict the lightest eigenvector with a mass $125.9$ GeV, and the choice is good for the behavior of $h\rightarrow\gamma\gamma$ and $h\rightarrow VV^*\;(V=Z,\;W)$ .\cite{ref11}

\begin{eqnarray}
&&
{B_4} = \frac{3}{2},{\kern 1pt} {\kern 1pt} {\kern 1pt} {\kern 1pt} {\kern 1pt} {\kern 1pt} {\kern 1pt} {\kern 1pt} {\kern 1pt} {\kern 1pt} {v_{{B_t}}} = \sqrt {v_B^2 + \bar v_B^2}  = 3{\rm{TeV}},\nonumber\\
 &&
\tan \beta  = \tan {\beta _B} = 2,\nonumber\\
 &&
{m_{{{\tilde U}_4}}} = {m_{{{\tilde Q}_5}}} = {m_{{{\tilde U}_5}}} = 1{\rm{TeV}},\nonumber\\
 &&
{A_{{u_4}}} = {A_{{u_5}}} = 500{\rm{GeV}},\nonumber\\
 &&
{A_{BU}} = 1{\rm{TeV}},{\kern 1pt} {\kern 1pt} {\kern 1pt} {\kern 1pt} {\kern 1pt} {\lambda _u} = 0.5,\nonumber\\
 &&
{Y_{{u_4}}} = 0.76{Y_t},{\kern 1pt} {\kern 1pt} {\kern 1pt} {\kern 1pt} {\kern 1pt} {Y_{{d_4}}} = 0.7{Y_b},{\kern 1pt} {\kern 1pt} \nonumber\\
 &&
{Y_{{u_5}}} = 0.7{Y_b},{\kern 1pt} {\kern 1pt} {\kern 1pt} {\kern 1pt} {\kern 1pt} {\kern 1pt} {\kern 1pt} {\kern 1pt} {Y_{{d_5}}} = 0.13{Y_t},{\kern 1pt} {\kern 1pt} {\kern 1pt} {\kern 1pt} {\kern 1pt} \nonumber\\
 &&
\mu  =  - 800{\rm{GeV}}{\kern 1pt} {\kern 1pt} {\kern 1pt} {\kern 1pt} {\kern 1pt} {\kern 1pt} {\kern 1pt}\nonumber\\
 &&
{B_X} = 500{\rm{GeV}},{\kern 1pt} {\kern 1pt} {\kern 1pt} {\kern
1pt} {\kern 1pt} {\mu _X} = 2{\rm{TeV}} ,\label{Higgs-EQ-1/3}
\end{eqnarray}

%%%%%%%%%%%%%%%%%%%%%%%%%%%%%%%%%%%%%%%%%%%%%%%%%%%%%%%%%%%%%%%%%%%%%%%%%%%%%%%%%%%%5
\begin{figure}\small
  \centering
   \includegraphics[width=6in]{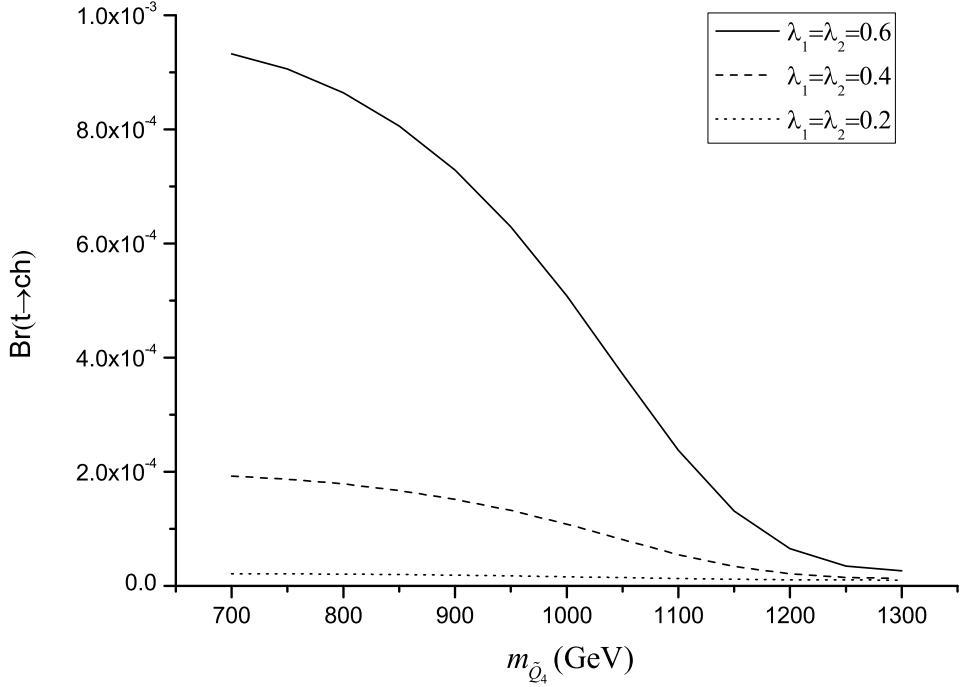}
     \caption{The branching ratio of  $t\rightarrow
ch$ varying with  $m_{\tilde{Q}_4}$}
    \label{fige3c}
\end{figure}
%%%%%%%%%%%%%%%%%%%%%%%%%%%%%%%%%%%%%%%%%%%%%%%%%%%%%%%%%%%%%%%%%%%%%%%%%%%%%%%%%%%%%%%%

Choosing $ {m_{{Z_B}}} = 1{\rm{TeV}}, {\mu _B} =
500{\rm{GeV}},{\lambda _Q} = 0.5,A_{BQ} = 1{\rm{TeV}}$. We plot in
Fig.3 the BRs of $t\rightarrow ch$ versus $m_{\tilde{Q}_4}$,
 the solid line ,dash line and dot line correspond to $\lambda_1=\lambda_2=
0.6,0.4,0.2$, respectively. We could see that the BRs decrease as
$m_{\tilde{Q}_4}$ runs from 700GeV to 1300GeV, and increase when
$\lambda_1=\lambda_2$ increase,  because $m_{\tilde{Q}_4}$ is the mass parameter of the exotic quarks, and $\lambda_1, \lambda_2$ proportional to the coupling coefficient. In addition, when
$m_{\tilde{Q}_4}\geq1100 $,  the BRs is tend to the results of MSSM.

%%%%%%%%%%%%%%%%%%%%%%%%%%%%%%%%%%%%%%%%%%%%%%%%%%%%%%%%%%%%%%%%%%%%%%%%%%%%%%%%%%%%5
\begin{figure}\small
  \centering
   \includegraphics[width=6in]{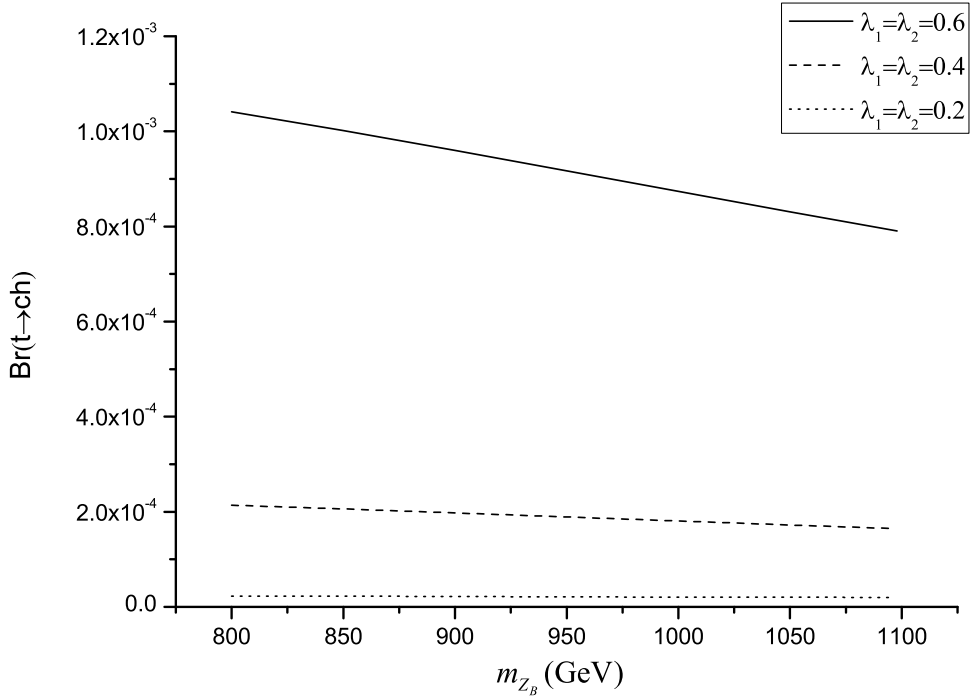}
     \caption{The branching ratio of  $t\rightarrow
ch$ versus $m_{Z_B}$}
    \label{fige3c}
\end{figure}
%%%%%%%%%%%%%%%%%%%%%%%%%%%%%%%%%%%%%%%%%%%%%%%%%%%%%%%%%%%%%%%%%%%%%%%%%%%%%%%%%%%%%%%%

In Fig. 4, we plot Br$(t\rightarrow ch)$ varying with $m_{Z_B}$.
Adopting $ m_{\tilde{Q}_4}=790{\rm{GeV}}, {\mu _B} =
500{\rm{GeV}},{\lambda _Q} = 0.5,A_{BQ} = 1{\rm{TeV}}$, and with
$\lambda_1=\lambda_2=0.6$(solid line),
$\lambda_1=\lambda_2=0.4$(dash line), $\lambda_1=\lambda_2=0.2$(dot
line). We could see that the BRs decrease as $m_{Z_B}$  runs from
800GeV to 1100GeV, since $m_{Z_B}$  contribute to the mass matrix of exotic squarks, and increase when $\lambda_1=\lambda_2$ increase.
And when $\lambda_1=\lambda_2=0.6,0.4$, Br$(t\rightarrow ch)$ is at the
order of $10^{-4}$, when $\lambda_1=\lambda_2=0.2$,
 Br$(t\rightarrow ch)$  is at the order of $10^{-5}$.

%%%%%%%%%%%%%%%%%%%%%%%%%%%%%%%%%%%%%%%%%%%%%%%%%%%%%%%%%%%%%%%%%%%%%%%%%%%%%%%%%%%%5
\begin{figure}\small
  \centering
   \includegraphics[width=6in]{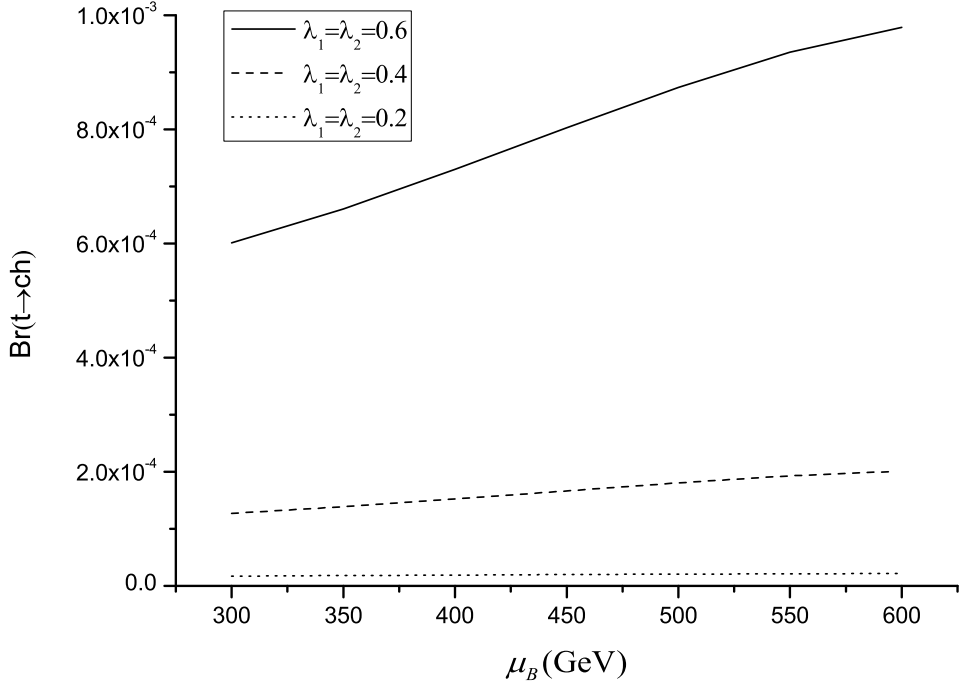}
     \caption{The branching ratio of  $t\rightarrow
ch$ as a function of $\mu_B$}
    \label{fige3c}
\end{figure}

%%%%%%%%%%%%%%%%%%%%%%%%%%%%%%%%%%%%%%%%%%%%%%%%%%%%%%%%%%%%%%%%%%%%%%%%%%%%%%%%%%%%%%%%

We assume $  m_{\tilde{Q}_4}=790{\rm{GeV}}, {m_{{Z_B}}} =
1{\rm{TeV}}, {\lambda _Q} = 0.5,A_{BQ} = 1{\rm{TeV}}$. We plot in
Fig.5 the BRs of $t\rightarrow ch$ versus $\mu_B$,
 the solid line ,dash line and dot line correspond to $\lambda_1=\lambda_2=
0.6,0.4,0.2$, respectively. We could see that the BRs increase as
${\mu_B}$ runs from 300GeV to 600GeV, since  $\mu_B$  inversely to the mass of exotic squarks,

%%%%%%%%%%%%%%%%%%%%%%%%%%%%%%%%%%%%%%%%%%%%%%%%%%%%%%%%%%%%%%%%%%%%%%%%%%%%%%%%%%%%5
\begin{figure}\small
  \centering
   \includegraphics[width=6in]{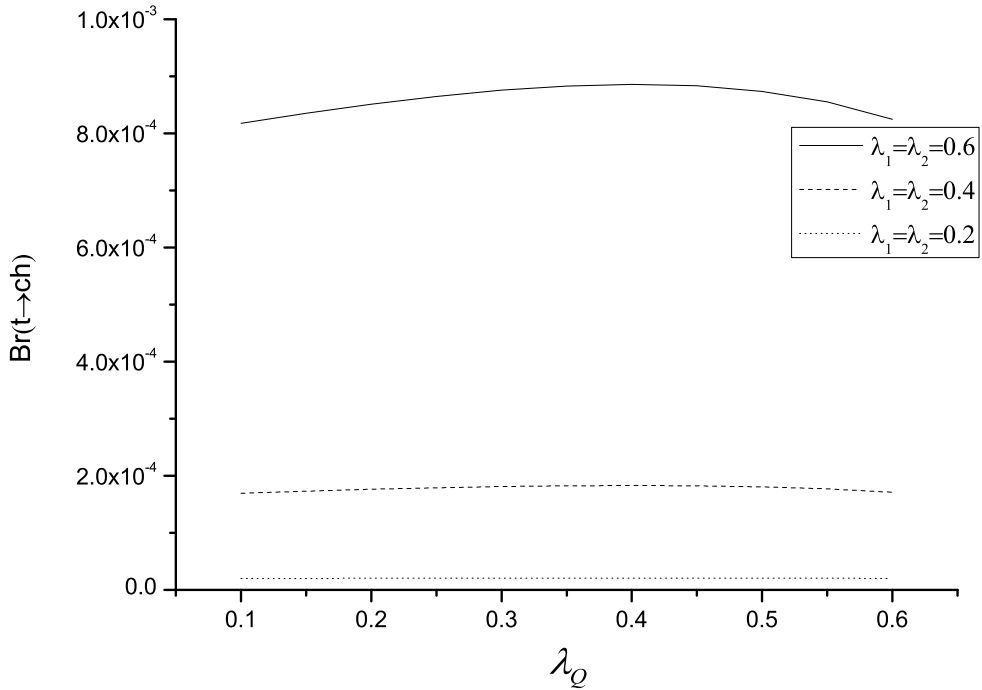}
     \caption{The branching ratio of  $t\rightarrow
ch$ varying with $\lambda_Q$}
    \label{fige3c}
\end{figure}
%%%%%%%%%%%%%%%%%%%%%%%%%%%%%%%%%%%%%%%%%%%%%%%%%%%%%%%%%%%%%%%%%%%%%%%%%%%%%%%%%%%%%%%%

Choosing ${m_{{{\tilde Q}_4}}} = 790{\rm{GeV}}, {m_{{Z_B}}} =
1{\rm{TeV}}, {\mu _B} = 500{\rm{GeV}},A_{BQ} = 1{\rm{TeV}}$, we draw
Br$(t\rightarrow ch)$  varying with $\lambda_Q$  in Fig.6
 for $\lambda_1=\lambda_2=0.6,0.4,0.2$ respectively.  We could see that the curve first increase and then decrease, but not
 significantly, since   $\lambda_Q$  contribute both to the mass of exotic squarks and the coupling coefficient,
.
%%%%%%%%%%%%%%%%%%%%%%%%%%%%%%%%%%%%%%%%%%%%%%%%%%%%%%%%%%%%%%%%%%%%%%%%%%%%%%%%%%%%5
\begin{figure}\small
  \centering
   \includegraphics[width=6in]{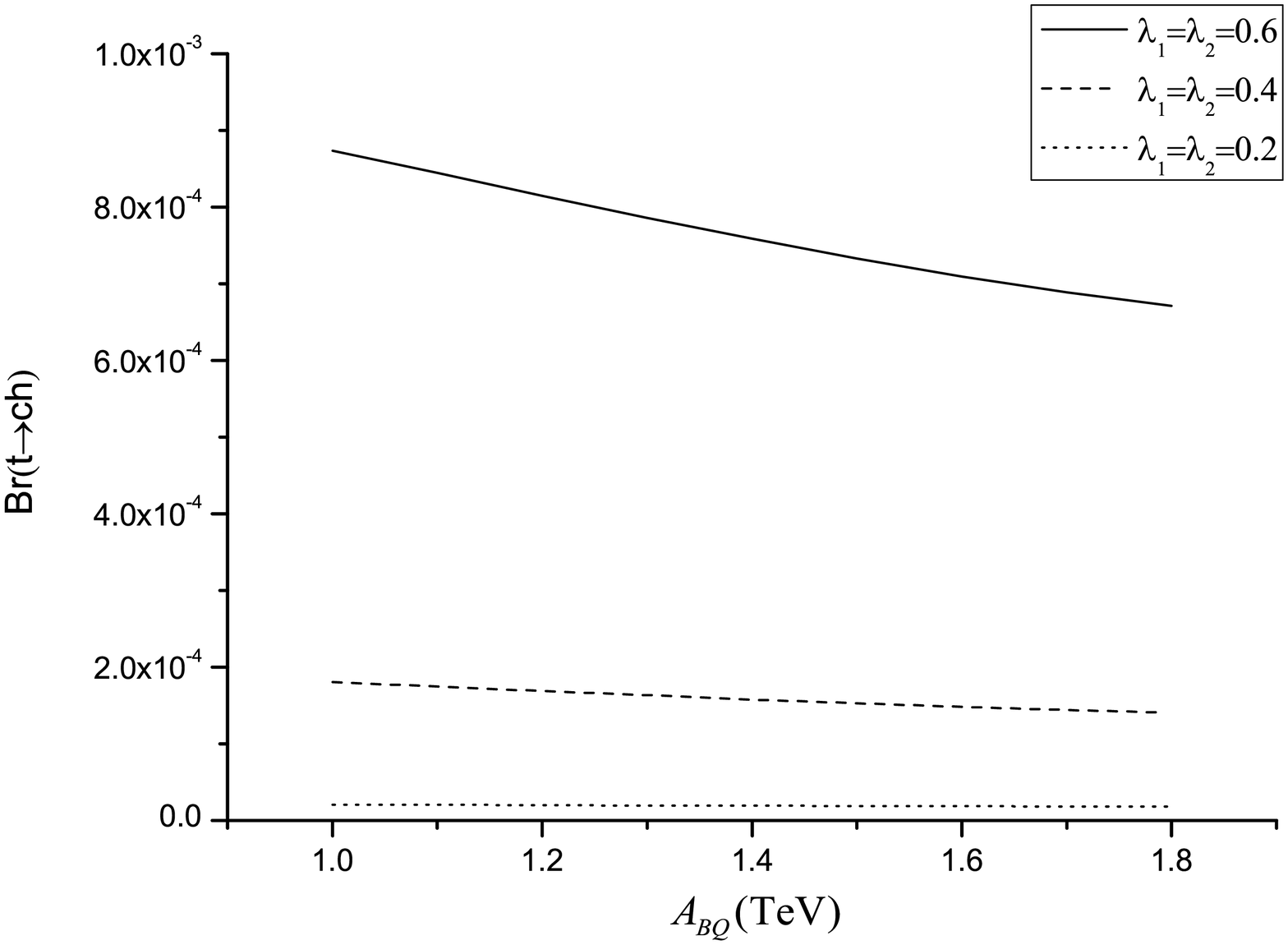}
     \caption{The branching ratio of  $t\rightarrow
ch$ versus $A_{BQ}$}
    \label{fige3c}
\end{figure}
%%%%%%%%%%%%%%%%%%%%%%%%%%%%%%%%%%%%%%%%%%%%%%%%%%%%%%%%%%%%%%%%%%%%%%%%%%%%%%%%%%%%%%%%

Taking ${m_{{{\tilde Q}_4}}} = 790{\rm{GeV}}, {m_{{Z_B}}} =
1{\rm{TeV}}, {\mu _B} = 500{\rm{GeV}},{\lambda _Q} = 0.5$,
 we show the Br$(t\rightarrow ch)$
varying with $A_{BQ}$ in Fig.7 for $\lambda_1=\lambda_2=0.6$(solid
line), $\lambda_1=\lambda_2=0.4$(dash line),
$\lambda_1=\lambda_2=0.2$(dot line), respectively. We could see that
the BRs decrease as $A_{BQ}$ runs from 1TeV to 1.8TeV, since $A_{BQ}$ contribute to the mass matrix of exotic squarks. And when
$\lambda_1=\lambda_2=0.6,0.4$, Br$(t\rightarrow ch)$ is at the order of
$10^{-4}$, when $\lambda_1=\lambda_2=0.2$,
 Br$(t\rightarrow ch)$  is at the order of $10^{-5}$.

\section{Summary \label{sec3}}
\indent\indent

The running LHC is a top-quark factory, and provides a great
opportunity to seek out top-quark decays. And it is showed that the channel $t\rightarrow ch$ could be
detectable  reaching a sensitivity level of Br$(t\rightarrow ch)\sim 5\times 10^{-5}$
\cite{ref14,ref15}.
But the fact is that the branching ratio of the process is so small in the SM\cite{ref6},
which is $\rm{Br}(t\rightarrow ch)\sim10^{-13}$, so it is too small to be measurable in the near future.

In this work, we study the rare top decay to a 125GeV Higgs in the framework of the BLMSSM.
Adopting reasonable assumptions on the parameter space, we present the radiative
correction to the process in BLMSSM, and draw some  curves between the BRs and new physics parameters.
We get the branching ratio of $t\rightarrow ch$ can reach $10^{-3}$, so this process could be detected in near future at LHC.

In addition, the author of \cite{ref16} yields an estimated upper limit of Br$(t\rightarrow ch) < 2.7\%$ for a Higgs boson mass of 125GeV, by combining the CMS results from a number of exclusive three- and four-lepton search channels.   And the ATLAS find the limit of Br$(t\rightarrow ch) < 0.83\%$ at $95\%$ C.L. by searching for $t\rightarrow ch$, with $h\rightarrow\gamma\gamma$, in $\bar{t}t$ events.\cite{ref17,ref18}  And our numerical evaluations indicates the BRs is highly dependent upon the parameters $\lambda_{1,2}$, the sensitive
parameters can make the contribution to  Br$(t\rightarrow ch)$ sizeable. Considering the experiment upper bounds  from CMS and ATLAS, the parameters $\lambda_{1,2}$ should not be too large under our assumptions of the parameter space.

As we could see above, the  $t\rightarrow ch$ process may be found in near future, and further
more constraints of BLMSSM can be obtained from more precise determinations.

\begin{acknowledgments}
\indent\indent The work has been supported by the National Natural Science Foundation of China (NNSFC)
with Grant No. 11275036, No. 11047002, the open project of State Key Laboratory of
Mathematics-Mechanization with Grant No. Y3KF311CJ1, the Natural Science Foundation of Hebei province with Grant No. A2013201277, and Natural Science Fund of Hebei
University with Grant No. 2011JQ05, No. 2012-242.
\end{acknowledgments}

\end{document}